\documentclass[journal]{IEEEtran}
\usepackage{url,amsmath,graphicx}
 \usepackage{stfloats}
 \usepackage{eqparbox}
  \usepackage{array}
  \usepackage{fixltx2e}
  \usepackage{stfloats}
   \usepackage{cite}
   \usepackage{url}
\usepackage{fancyhdr}
\usepackage{cite}
\usepackage{graphicx}
\usepackage{psfrag}
\usepackage{subfigure}
\usepackage{url}
\usepackage{stfloats}
\usepackage{amsmath}
\usepackage{array}
\usepackage{fancyhdr}
\usepackage{epsfig}
\usepackage{amssymb}
\usepackage{color}

\def\he{{\mathcal{H}}}

\begin{document}
\title{ HetNets Coverage Modeling and Analysis Over Fox's $\mathcal{H}$-Fading Channels  }
\author{ Im\`ene~Trigui and Sofi\`ene~Affes\\
INRS-EMT, 800, de la Gaucheti\`{e}re Ouest, Bureau
6900, Montr\'{e}al, H5A 1K6, Qc, Canada. \\
\{itrigui, affes\}@emt.inrs.ca
}
 \maketitle

\begin{abstract}
This paper embodies the Fox's ${\mathcal H}$-transform theory into a unifying modeling and analysis of HetNets.  The proposed framework has the potential, due to  the Fox's ${\mathcal H}$-functions versatility, of significantly simplifying the cumbersome analysis
 and representation of  cellular coverage, while subsuming  those previously derived for all known simple and composite
fading models. The paper reveals important
insights into the practice of densification in conjunction
with signal-to-noise plus interference (SINR) thresholds and path-loss models.\let\thefootnote\relax\footnotetext{Work supported by the Discovery Grants Program of NSERC and a Discovery Accelerator Supplement (DAS) Award from NSERC. It was first submitted to IEEE Communications Letters on 12 June 2018.}

\end{abstract}

\section{Introduction}
	\IEEEPARstart{C}hiefly urged by the occurring mobile data deluge, a radical design make-over of cellular systems advocating heterogenous
cellular networks (HetNets) is crucial and thus an active research trend \cite{and3}-\!\!\cite{gupta}. The random space pattern of HetNets has been extensively reproduced and analyzed  trough  stochastic geometry over different  fading channels such as Rayleigh \cite{and}, Nakagami-$m$ \cite{sir}, Weibull \cite{trigui1} and  $\alpha$-$\mu$  \cite{trigui}.

Besides subsuming most of these fading models,  the Fox's $\mathcal H$-distribution is currently
being touted  for its  high flexibility to adapt different
fading behaviors pertaining to emerging new wireless applications, e.g., device-to-device (D2D) and intervehicular communications,  wireless body area networks,  and  millimeterwave (mmWave) communications \cite{cotton}.
Despite several studies on its  applicability in evaluating various wireless communication (\!\cite{yusuf} and references therein), the Fox's $\mathcal H$-distribution has thus far  not found its way into stochastic geometry-based cellular communications as a possible fading distribution. Yet, resorting to the most comprehensive treatments of the
subject \cite{and}-\!\!\cite{d2dsku}, a general analytic solution  for Fox's $\mathcal H$ fading seems unlikely, if not impossible.
Indeed, besides being  simple special
cases, these treatments  rely   on approximating the fading distribution
 (e.g., integer fading parameter-based power series \cite{sir}, \cite{gupta}, and Laguerre polynomial series in \cite{d2dsku}) which hamper their generality and exactness. Moreover,  these treatments  usually entail computationally expensive Laplace
generation functional evaluation lending the solution approach itself
complicated and more importantly non applicable to the generalized Fox's $\mathcal H$ fading.

To the best of the authors' knowledge, no work has ever been found to analyze the coverage of HetNets over the
general Fox's $\mathcal H$ fading channels. The main contributions of
this letter are as follows:
\begin{itemize}
  \item Novel exact and closed-form expressions are
derived for the coverage of HetNets over  Fox's $\mathcal H$-fading under both range expansion as well
as max-SIR cell association (CA) rules. Our  analysis procedure  and coverage formulations are given in unified and
tractable mathematical fashion thereby serving  as a useful tool
to validate and compare the special cases of Fox's $\mathcal H$-fading channels.
  \item  Some useful insights regarding the practice of densification of HetNets in conjunction
with path-loss model are also provided
through the asymptotic coverage analysis.
  \item The derived results enable to evaluate the impacts
of physical channel and network dynamics such as  fading parameters, density of
BSs, SINR thresholds, and path-loss model on  coverage performance.
\end{itemize}


 \section{Channel and Network  Models}
 \subsection{The Fox's $\mathcal{H}$ Channel Model}
Consider a wireless communication link over a  fading
channel where the power gain is distributed according to the Fox's $\mathcal{H}$ $\{{\cal O}, {\mathcal P}\}$ distribution
 with order sequence ${\mathcal O}= (m,n,p,q)$,   parameter sequence ${\mathcal P}= (\kappa, c, a, b, A, B)$, and probability density function (PDF)
\begin{equation}
f_{H}(x)= {\mathcal \kappa} {\mathcal H}_{p,q}^{m,n}\left[ c x \left|
\begin{array}{ccc} (a_i, A_j)_p \\ (b_i,B_j)_q \end{array}\right. \right], \quad x\geq0,
\label{eq1}
\end{equation}
where $c$ and $\kappa$ are constants, and  $(x_j , y_j )_l$ is a shorthand notation  for $(x_1, y_1),..., (x_l, y_l)$.
 Hereafter, for notational simplicity, we denote the right-hand side of
(\ref{eq1}) by ${\mathcal H}_{p,q}^{m,n}(x;{\cal P})$.

\subsection{Special cases}
A  Fox's H-function PDF considers homogeneous
radio propagation conditions and captures composite effects of multipath
fading and shadowing, subsuming
large variety of extremely important or generalized fading distributions
 used in wireless communications as $\alpha$-$\mu$\footnote{The $\alpha$-$\mu$ distributions can be attributed to exponential, one-sided
Gaussian, Rayleigh, Nakagami-m, Weibull and Gamma fading distributions
by assigning specific values for $\alpha$ and $\mu$.},  $N$-Nakagami-$m$, (generalized) ${\cal K}$-fading,
and Weibull/gamma fading , the Fisher-Snedecor
F-S $\cal F$, and EGK, as shown
in Table. I (\!\cite{yusuf}, \cite{aloui} and references therein). Furthermore, the Fox's H-function distribution provides enough flexibility  to account for disparate signal
	propagation mechanisms and  well-fitted to measurement data collected in
diverse propagation environments having different parameters.
\begin{table}\caption{Special
cases of Fox's H-function distribution}
\begin{equation}
\begin{array}{cc}\\
   \!\!\! f_{H}(x)\sim\mathcal{H}\{{\cal O}, {\mathcal P}\} \\ \hline
  \begin{array}{c}\alpha-\mu \\
 \end{array} &\\ \begin{array}{c}
                 {\mathcal O}_{\alpha-\mu}= (1,0,0,1) \\
               \!\!\!  \!\!\!{\mathcal P}_{\alpha-\mu}=\left(\frac{\Gamma(\mu+\frac{1}{\alpha})}{\Gamma(\mu)^{2}}, \kappa \Gamma(\mu), -, \mu-\frac{1}{\alpha}, -, \frac{1}{\alpha}\right)
               \end{array}\\  \hline\\
\textbf{F-S }{\cal F} & \\  \begin{array}{c}{\mathcal O}_{ \textbf{F-S }{\cal F}}=(1,1,1,1) \\  \!\!\! \!\!\! {\mathcal P}_{ \textbf{F-S }{\cal F}}=\left(\frac{c}{\Gamma(m)\Gamma(m_s)}, \frac{m}{m_s}, -m_s, m-1, 1, 1\right)\end{array}\\  \hline\\
\textbf{EGK} & \\\begin{array}{c}{\mathcal O}_{ \textbf{EGK}}=(2,0,0,2)\\  \!\!\! \!\!\! {\mathcal P}_{\textbf{EGK}}=\left(\frac{\beta \beta_s}{\Gamma(m)\Gamma(\kappa)},\beta \beta_s, -, (m-\frac{1}{\zeta}, \kappa_s-\frac{1}{\zeta}), -, (\frac{1}{\zeta},\frac{1}{\zeta})\right)\\ \text{where}~ \beta=\frac{\Gamma(m+\frac{1}{\zeta})}{\Gamma(m)} ~\text{and}~ \beta_s=\frac{\Gamma(\kappa_s+\frac{1}{\zeta})}{\Gamma(m)} \end{array}\\ \hline \end{array}
\nonumber
\end{equation}
\end{table}

 \subsection{Network Model}
Consider the downlink of a ${\cal M}$-tier HetNet. Each tier
 is specified by the tuple $(\lambda_i, P_i, \beta_i, \{{\cal O}_i, {\mathcal P}_i\} ), i\in\{1,\ldots, {\cal M}\}$,  indicating
the BS spatial density, transmission power, target SINR threshold, and the order and parameters sequences of the $\mathcal{H}$-fading, respectively.
The BSs in the $i$-th tier  are spatially distributed as a homogenous Poisson
point process (PPP) $\Phi_i \in \mathbb{R}^{2}$  with density $\lambda_i$.
Let  $H_{x_i}$ be the  channel power
gain between BS $x_i \in \Phi_i$
to be distributed according to the Fox's $\mathcal{H}$-distribution $\{{\cal O}_i, {\mathcal P}_i\}$. Furthermore,  we denote $L(\|x\|)$  the path-loss function and
\begin{equation}
{\cal I}=\sum_{i\in {\cal M}}\sum_{x_i\in \Phi_i/ x_k}P_i L(\|x_i\|)H_{x_i},
\end{equation}
the aggregate interference at a typical receiver, assuming that its serving BS belongs
to the $k$-th tier.  The SINR at the typical receiver can then  be formulated as\vspace{-0.1cm}
\begin{equation}
\text{SINR}^{m}_{x_k}=\frac{P_k L(\|x_k\|)H_{x_k}}{{\cal I}+\sigma^{2}_k},
\end{equation}
where $\sigma^{2}_k$  is the thermal noise power associated with the $k$-th tier, and the parameter
$m\in\{{\cal U}, {\cal B}\}$ where i) $m={\cal U}$  stands for the unbounded path-loss scenario,  i.e., $L(\|x\|)=\|x\|^{-\alpha}$ where $\alpha$ is the path-loss exponent and
 ii) $m={\cal B}$ uses the bounded path-loss model, i.e.,  $L(\|x\|)=(1+\|x\|)^{-\alpha}$.

\section{Fox's $\mathcal{H}$ Modeling of Coverage}

\subsection{RSS Cell Association}
 Let the
typical user be associated with the BS that provides the maximum radio signal strength (RSS). This implies that the typical user is then in coverage
if the set $ {\cal A}^{m} = \left\{\exists i \in {\cal M}: i=\arg\max_{j\in {\cal M}, x\in \Phi_j} P_j L(\|x_j\|); \text{SINR}^{m}_{x_i}\geq \beta_i\right\}$
 is not empty.  Let us denote $r_k = \|x_k\|$ and define the coverage probability by ${\cal C}^{m} = P\{{\cal A}^{m}\neq \varnothing\}$.
${\cal C}^{m}\triangleq\sum^{{\cal M}}_{k=1} {\cal \theta}_k \mathcal{E}_{r_k}\{{\cal C}^{m}(r_k)\}$ where ${\cal \theta}_k= \frac{\lambda_k}{\sum_{j\in {\cal M}}\lambda_j \widetilde{P}_j^{\delta}}$.

\textit{Proposition 1:}
 The  average coverage probability Fox's $\mathcal{H}$-fading with an unbounded path loss model  is given by
\begin{eqnarray}
\!\!\!\!\!\!\!{\cal C}^{\cal U}\!\!\!&=& \!\!\!\pi \delta \sum^{{\cal M}}_{k=1} \lambda_k \left(\frac{P_k}{\sigma_k^{2}}\right)^{\delta}\int_{0}^{\infty}\frac{1}{\xi^{2+\delta}}{\mathcal H}_{q,p+1}^{n,m}\left(\xi, {\cal P}^{k}_{\cal U}\right)\nonumber\\ &&  \!\!\!\!\!\!\!\!\!\!\!\!\!\!\!\!\!\!\!\!\!\!\!\!{\mathcal  H}_{1,1}^{1,1}\!\!\left(\!\!\frac{\left(\frac{P_k}{\sigma^{2}_k}\right)^{\delta}}{\xi^{\delta}}\!\!\sum_{j\in {\cal M}}\!\!\! \pi \lambda_j \widetilde{P}^{\delta}_j \!\left(1\!+\!\delta \xi {\mathcal H}_{q+2,p+3}^{n+1,m+2}\!\!\left(\xi, {\cal P}^{{\cal I}}_{\cal U}\right)\!\right)\!,\!{\cal P}_{\delta}\!\!\right)\!d\xi,
\label{cb}
\end{eqnarray}
where $ {\cal P}^{k}_{\cal U}=\!\!\left(\kappa {\cal \beta}_k, \frac{1}{c {\cal \beta}_k}, 1\!-\!b, (1\!-\!a, 1), \mathcal{B}, (A,1)\right),
$ and ${\cal P}^{{\cal I}}_{\cal U}=\bigg(\frac{\kappa}{c^{2}}, \frac{1}{c}, (1\!-\!b\!-\!2B,0,\delta),(0, 1\!-\!a\!-\!2A, -1,\delta\!-\!1),$ $ (\mathcal{B},1,1), (1,A,1,1)\bigg)$, and
${\cal P}_{\delta}=\left(1,1, 1-\delta, 0, \delta, 1\right)$.

\textit{Proof:} See Appendix A.\\ The new fundamental SINR distribution disclosed in Proposition 1 provides an exact  and numerically inexpensive unifying tool for coverage analysis in a variety of extremely important fading distributions (see \cite[Table I]{aloui}). In some particular cases, the obtained formulas reduces to previously well-known major results in the literature\footnote{For instance, the Fox's $\mathcal{H}$ distribution with  ${\cal O}=(1,0,1,0)$ and $P=(1, 1, 0,1,0,1)$ reduces to Rayleigh fading for which indeed (\ref{cb}) matches the classical results readily available in the literature \cite[Eq. (14)]{trigui1}, \cite[Theorem 1]{and}.}  \cite{and},\!\cite{dilhon},\! \cite{sir}.

\textit{Corollary 1 (HetNets densification in Fox's $\mathcal{H}$-fading with unbounded path-loss model):} The average coverage of ultra-dense networks with RSS under unbounded path-loss scales as
\begin{eqnarray}
\lim_{\lambda\rightarrow\infty} {\cal C}^{\cal U}= \nonumber \\ && \!\!\!\!\!\!\!\!\!\!\!\!\!\!\!\!\!\!\!\!\!\!\!\!\!\!\!\!\!\!\!\sum^{{\cal M}}_{k=1}\!\int_{0}^{\infty}\frac{{\mathcal H}_{q,p+1}^{n,m}\left(\xi, {\cal P}^{k}_{\cal U}\right)d\xi}{\xi^{2}\sum_{j\in {\cal M}}\! \widetilde{P}^{\delta}_j\! \left(1\!+\!\delta \xi {\mathcal H}_{q+2,p+3}^{n+1,m+2}\!\!\left(\xi, {\cal P}^{{\cal I}}_{\cal U}\right)\right)}.
\label{cul}
\end{eqnarray}

\textit{Proof:}
Recall that the asymptotic expansion of the Fox's $\mathcal{H}$-function near $x = \infty$ given by \cite[Eq. (1.5.9)]{kilbas}
\begin{equation}
{\mathcal H}_{p,q}^{m,n}(x;{\cal P})\underset{x\rightarrow\infty}{\approx} \kappa \eta x^{d},
\label{Has1}
\end{equation}
where $d= \max \left(\frac{a_i-1}{A_i}\right), i=1,\ldots,n$ and $\eta$ is calculated as in  \cite[Eq. (1.5.10)]{kilbas}.
Applying (\ref{Has1}) to (\ref{cb}) when $\lambda_k=\lambda\rightarrow\infty, k=1,\ldots, {\cal M} $,  yields the result after recognizing that $d=-1$ and $\eta=\frac{1}{\delta}$.

Corollary 1 shows how the singularity in the unbounded model
can  affect the accountability of the conducted analysis,  since the coverage intensity-invariance property of ultra-dense  HetNets still holds under the Fox's $\mathcal{H}$-fading.

\textit{Proposition 2:} The  coverage probability over Fox's $\mathcal{H}$-fading with a bounded path-loss model for a receiver connecting to the $k$-th tier BS  located at $x_k$  is given by
\begin{eqnarray}
{\cal C}^{\cal B}(r_k)\!\!&=&\!\!\int_{0}^{\infty}\frac{1}{\xi^{2}}{\mathcal H}_{q,p+1}^{n,m}\left(\xi, {\cal P}^{k}_{\cal B}\right)\exp\Bigg(-\frac{\sigma^{2}}{P_k}\xi (1+r_k)^{\alpha}\nonumber\\ &&\!\!\!\!\! \!\!\!\!\!\!\!-\sum_{j\in {\cal M}} \pi \lambda_j \widetilde{P}^{\delta}_j\delta \xi \Bigg((1+r_k)^{2}{\mathcal H}_{q+2,p+3}^{n+1,m+2}\left(\xi, {\cal P}^{1,{\cal I}}_{\cal B}\right)\nonumber \\ && -(1+r_k){\mathcal H}_{q+2,p+3}^{n+1,m+2}\left(\xi, {\cal P}^{2,{\cal I}}_{\cal B}\Bigg)\right)\Bigg)d\xi,
\label{mgf}
\end{eqnarray} where ${\cal P}^{k}_{\cal B}={\cal P}^{k}_{\cal U}$,  ${\cal P}^{1,{\cal I}}_{\cal B}= \bigg(\frac{\kappa}{c^{2}}, \frac{1}{c}, (1-b-2B,0,\delta),(0, 1-a-2A, -1,\delta-1), (\mathcal{B},1,1), (1,A,1,1)\bigg)$, and ${\cal P}^{2,{\cal I}}_{\cal B}=\bigg(\frac{\kappa}{c^{2}}, \frac{1}{c}, \left(1\!-\!b\!-\!2B,0,\frac{\delta}{2}\right),\left(0, 1\!-\!a\!-\!2A, -1,\frac{\delta}{2}\!-\!1\right),$ $ (\mathcal{B},1,1), (1,A,1,1)\bigg)$.

\textit{Proof:} Appendix B.

\textit{Corollary 2:} In interference-limited HetNets, the average coverage probability over Fox's $\mathcal{H}$-fading with a bounded path-loss model is obtained as
\begin{eqnarray}
\!\!\!\!{\cal C}^{\cal B}&=&\sum_{k\in {\cal M}}\lambda_k \int_{0}^{\infty}\!\!\!\frac{e^{-\sum_{j\in {\cal M}} \pi \lambda_j \widetilde{P}^{\delta}_j\delta \xi (\Psi_1-\Psi_2)}}{\xi^{2}\sum_{j\in {\cal M}} \pi \lambda_j \widetilde{P}^{\delta}_j\delta \xi (\Psi_1+\Psi_2)}\nonumber\\ &&\!\!\!\!\! \!\!\!\!\!\!\!\!\! \!\!\!\!\!\!\!\!\!\!\!{\mathcal H}_{q,p+1}^{n,m}\!\!\left(\xi, {\cal P}^{k}_{\cal B}\right){\mathcal H}_{1,1}^{1,1}\!\!\left(\!\!\frac{\sum_{j\in {\cal M}} \lambda_j \widetilde{P}^{\delta}_j\left(1\!+\!\delta \xi \Psi_1\right)}{\sum_{j\in {\cal M}} \lambda_j \widetilde{P}^{\delta}_j\delta \xi(2 \Psi_1\!+\!\Psi_2)}, {\cal \widetilde{P}}_{\delta}\!\!\right)\!d\xi,
\label{mgf}
\end{eqnarray}
where $\Psi_x={\mathcal H}_{q+2,p+3}^{n+1,m+2}\left(\xi, {\cal P}^{x,{\cal I}}_{\cal B}\right), x\in\{1,2\}$ and ${\cal \widetilde{P}}_{\delta}=\left(1,1, -1, 0, 2, 1\right)$.

\textit{Proof:} Since the BS density is typically quite high in HetNets, the interference power easily dominates
thermal noise. Therefore, thermal noise can often  be neglected i.e. $\sigma_k^{2}=0, k=\{1,\ldots, {\cal M}\}$. Then  the result follows along the same lines as in (\ref{cb}) after expanding $(1+r_k)^{2}$.

\textit{Corollary 3 (HetNets densification in Fox's $\mathcal{H}$-fading with bounded path-loss model):} The average coverage of ultra-dense networks with with RSS CA and under bounded path-loss  scales as
\begin{eqnarray}
\!\!\!\lim_{\lambda\rightarrow\infty} {\cal C}^{\cal B}\!\!\!\!&=&\!\!\!\!\!\!\sum_{k\in {\cal M}}\int_{0}^{\infty}\frac{e^{-\lambda \sum_{j\in {\cal M}} \pi  \widetilde{P}^{\delta}_j\delta \xi (\Psi_1-\Psi_2)}{\mathcal H}_{q,p+1}^{n,m}\left(\xi, {\cal P}^{k}_{\cal B}\right)}{\xi^{2}\sum_{j\in {\cal M}} \pi  \widetilde{P}^{\delta}_j\delta \xi (\Psi_1+\Psi_2)}\nonumber\\ && \!\!\!\!\!\!\!\!\nonumber\\ &&\!\!\!\!\!\!\!\!{\mathcal H}_{1,1}^{1,1}\left(\frac{\sum_{j\in {\cal M}} \widetilde{P}^{\delta}_j(1+\delta \xi \Psi_1)}{\sum_{j\in {\cal M}}  \widetilde{P}^{\delta}_j\delta \xi(2 \Psi_1+\Psi_2)}, {\cal P}_{\delta}\right)d\xi.
\end{eqnarray}
Contrary to what the  standard
unbounded path-loss function  predicts, the coverage
probability under bounded path-loss function
scales with $e^{-\lambda}$ and approaches zero with increasing $\lambda$ for general values of $\delta$. Recently, the authors in \cite{dens} revealed that the same can be spotted
in a single-tier cellular network over Rayleigh fading. Due to the complexity of the bounded  model, its impact was only understood through approximations in  \cite{sir},  yet merely for fading scenarios  with integer parameters. In this paper, ultra densification is scrutinized in HetNets over the Fox's $\mathcal{H}$-fading, which is to the best of our
 knowledge totally new.

\subsection{ Max-SINR Cell Association}
Under the max-SINR CA rule, the typical user is in coverage
if the set $ {\cal A}^{m} = \left\{\exists i \in {\cal M}; \underset{x_i\in \Phi_i} \max \text{SINR}^{m}_{x_i}\geq {\cal \beta}_i\right\}$ is not empty \cite{sir}. Then the average coverage
probability follows from  \cite[Lemma 1]{dilhon} as
\begin{equation}
{\cal C}^{m}= 2\pi \sum_{k\in {\cal M}}\lambda_k\int_{0}^{\infty}\!\!r_k {\cal C}^{ m}(r_k)d_{r_k}, ~~ m\in\{\cal U, \cal B\}.
\label{cm}
\end{equation}

\textit{Proposition 3:}  The average coverage probability in Fox's-${\mathcal H}$ fading is
\begin{eqnarray}
\!\!\!{\cal C}^{\cal U}=\sum_{k\in {\cal M}}\frac{\lambda_k }{  \beta_k^{\delta}\Gamma(1+\delta)}~{\mathcal H}_{1,1}^{1,1}\left(\frac{P_k}{\sigma_k^{2}};\tilde{{\cal P}}_k \right)\Lambda^{m,n}_{p,q},
\label{outcu}
\end{eqnarray}
where $\tilde{{\cal P}}_k=\Bigg(\frac{\pi}{\Delta}, \Delta^{\frac{1}{\delta}}, 1, 1,1, \frac{1}{\delta}\Bigg)$,
with $\Delta=\sum_{j\in {\cal M}}\pi\lambda_j \widetilde{P}_j^{\delta} \Gamma(1-\delta)\Lambda^{m,n}_{p,q}$, and
\begin{eqnarray}
\Lambda^{m,n}_{p,q}&=&\frac{\kappa}{c^{\delta+1}}\frac{\prod_{j=1}^{m}\Gamma\left(b_j+(1+\delta)B_j\right)
}{\prod_{j=m+1}^{p}\Gamma\left(1-b_j-(1+\delta)B_j\right)}\nonumber\\ && \times \frac{\prod_{j=1}^{n}\Gamma\left(1-a_j-(1+\delta)A_j\right)}{\prod_{j=n+1}^{p}\Gamma\left(a_j+(1+\delta)A_j\right)},
 \end{eqnarray}

\textit{Proof:} See Appendix C.

Notice that in contrary to \cite{sir}, \cite{d2dsku} our analysis procedure  and coverage formulations are not submissive to any
restrictive assumptions or approximation.  Indeed the coverage formulas in (\ref{outcu}) is generally enough to cover any fading distribution by simply tuning the Fox's H-function parameters, countless in number.
Remarkably,  this is the first unified and closed-form coverage formulas under generalized fading with Fox's H-function PDF.

\textit{Corollary 4:}   In an interference-limited network, the average coverage
probability simplifies from (\ref{outcu}) as
\begin{eqnarray}
\!\!\!{\cal C}^{\cal U}\!\!\!\!&=& \!\!\!\!\!\frac{\pi}{C(\delta)}  \sum_{k\in {\cal M}}\frac{\lambda_k \beta_k^{-\delta}\Lambda^{m,n}_{p,q} }{\sum_{j\in {\cal M}}\lambda_j \widetilde{P}_j^{\delta} \Lambda^{m,n}_{p,q}},
\label{cuas}
\end{eqnarray}
where $C(\delta)=\pi^{2}\delta~{\rm csc}(\pi \delta)$.

\textit{Proof:}
When $\sigma^{2}_k\simeq0$ it holds that in (\ref{outcu})
\begin{equation}
{\mathcal H}_{1,1}^{1,1}\left(\frac{P_k}{\sigma_k^{2}};\tilde{{\cal P}}_k \right)\underset{\sigma_k^{2}\simeq0}{\approx}\frac{\pi}{\Delta},
\end{equation}
thereby yielding the desired result.

 From  (\ref{cuas}), it follows that, unless $\{{\cal O}_i,{\cal P}_i\}\neq \{{\cal O}_j,{\cal P}_j\}$, $\forall \{i, j\}= 1,\ldots, {\cal M}$ (non identically distributed tiers), the coverage probability  is not affected by fading in an interference-limited network.  Remarkably, (\ref{cuas}) is instrumental in evaluating the impacts  of the number
of tiers or their relative densities,  transmit powers, and  target SINR  over generalized fading
scenarios. Strictly speaking, this result fills the gap of lacking exact, unified and simple coverage
expression over those fading channels.

Accommodating the closed-form expressions for coverage
performance  in the corresponding entries in Table I,
directly yields the results. After some
simple algebraic manipulations, one can observe the obtained
results herein are identically consistent with the existing works. For instance, under  $\alpha$-$\mu$ fading we obtain

\begin{eqnarray}
{\cal C}^{{\cal U}, \alpha-\mu}&=&\frac{\pi}{C(\delta)} \nonumber \\ && \!\!\!\!\!\!\!\!\!\!\!\!\!\!\!\!\sum_{k\in {\cal M}}\frac{\lambda_k \beta_k^{-\delta}\frac{\Gamma(\mu_k)^{\delta-1}}{\Gamma\left(\mu_k+\frac{1}{\alpha_k}\right)^{\delta}} \Gamma\left(\mu_k+\frac{\delta}{\alpha_k}\right) }{\sum_{j\in {\cal M}}\lambda_j \widetilde{P}_j^{\delta} \frac{\Gamma(\mu_j)^{\delta-1}}{\Gamma\left(\mu_j+\frac{1}{\alpha_j}\right)^{\delta}} \Gamma\left(\mu_j+\frac{\delta}{\alpha_j}\right)}.
\label{outalp}
\end{eqnarray}
Notice that when $\mu=m$ and $\alpha=1$, (\ref{outalp}) boils down to the coverage of HetNets under  arbitrary Nakagami-$m$ fading. The latter has been tackled in closed-form only when $m$ is an integer \cite{sir}, while the general case has been the subject of several ad hoc approximations \cite[Proposition 1]{sir}, \cite[Corollary 1]{dilhon}.

\textit{Proposition 4:} The average  coverage
probability of max-SINR CA with a bounded path-loss model over Fox's-${\mathcal H}$ fading is obtained as
\begin{eqnarray}
{\cal C}^{\cal B}\!\!\!&=&\!\!\!\!  2\pi \!\!\sum_{k\in {\cal M}}\!\lambda_k\!\!\int_{0}^{\infty}\!\!\frac{1 }{\xi^{2}}{\mathcal H}_{q,p+1}^{n,m}\left(\xi; {\cal P}^{k}_{\cal B}\right)\int_{0}^{\infty}\!\!\!\!\!r_k \exp\!\bigg(\!\!-(1\!+\!r_k)^{\alpha}\nonumber \\&& \!\!\!\!\!\!\!\!\!\!\!\!\!\!\!\!\!\!\!\!\Bigg(\!\!\frac{\sigma^{2}_k\xi}{P_k} \!+\!\!\!\sum_{j\in {\cal M}}\!\! \pi \lambda_j \delta \widetilde{P}^{\delta}_j \xi {\mathcal H}_{q+2,p+3}^{n+1,m+2}\!\left(\!\xi(1\!+\!r_k)^{\alpha}, {\cal P}^{1,{\cal I}}_{\cal B}\!\right)\!\!\!\Bigg)d_{r_k} d\xi.
\label{cm2}
\end{eqnarray}

\textit{Proof:} We obtain the result by proceeding along the same lines adopted in Appendix B in combination with (\ref{cm}).

\section{Numerical Results}
Fig.~1(a) shows the average coverage probability ${\cal C}^{ \cal U}$  under both RSS and max-SINR CA rules
vs. $\lambda_2$. It shows that for a small density of Tier $1$
 ($\lambda_1=10^{-4}$), densifying Tier $2$ steadily increases the coverage probability when $\beta_1>\beta_2$. Otherwise,  densificatoin of Tier $2$ always negatively affects the
coverage probability, even more dramatically when $\lambda_1$ is small.  Fig.~1(a) also shows that compared to the
max-SINR CA rule, the RSS scheme has much lower coverage
performance.

Fig.~1(b)  depicts the average coverage probability ${\cal C}^{ m}$, $m\in\{\cal U, \cal B\}$  with max-SINR CA.
It shows that the analysis is accurate and follows
the simulation trends.  Fig.~1(b) further validates
the explanations provided in section III regarding the impact of densification
on the coverage probability, as well as the impact of the
bounded model on the coverage probability versus the unbounded one. The former provides generally smaller coverage, particularly in
dense scenarios.

%
\begin{figure}[t]
\centering
\includegraphics[scale=0.3]{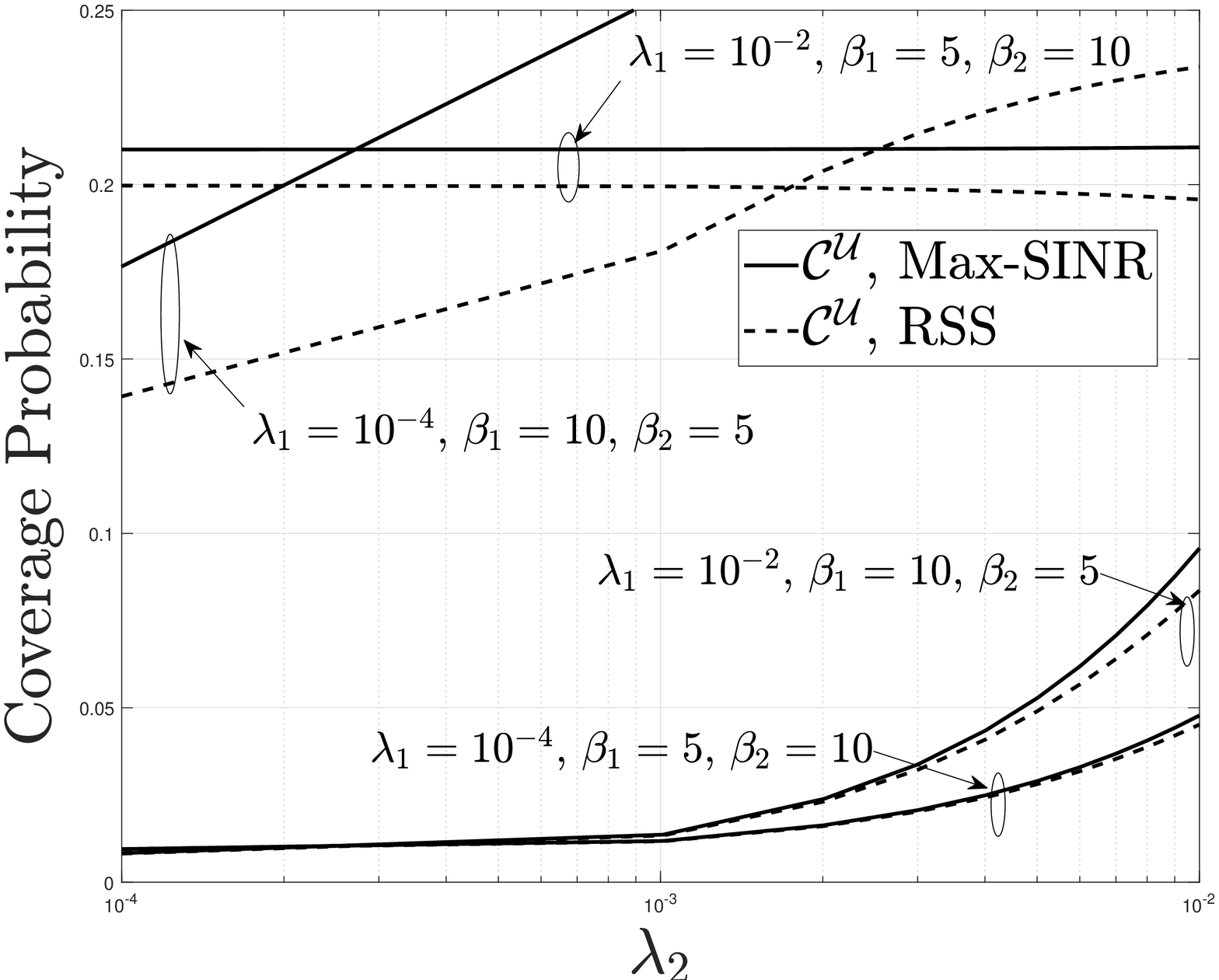}
\caption{Average coverage probability ${\cal C}^{ x}$  vs. $\lambda_2$  over $(3, 0, 0, 3)$-order  Fox's ${\mathcal H}$   for multipath fading  with ${\mathcal P}=(0.2, 5.5, -,(1.5, 0.4,4.5), - , \frac{1}{2}{\bf 1}_3)$. ${\cal M} = 2$, $\alpha=4$, $P_1 = 50$ W, $P_2 = 1$ W, and $\sigma_1^{2}=\sigma_2^{2}=10^{-6}$,  for: (a) both RSS and max-SINR CA schemes and $x =  \cal {U}$, and (b)  max-SINR CA and  $x \in  \{ \cal {U}, \cal {B}\}$..}
\end{figure}
\begin{figure}[t]
\centering
\includegraphics[scale=0.3]{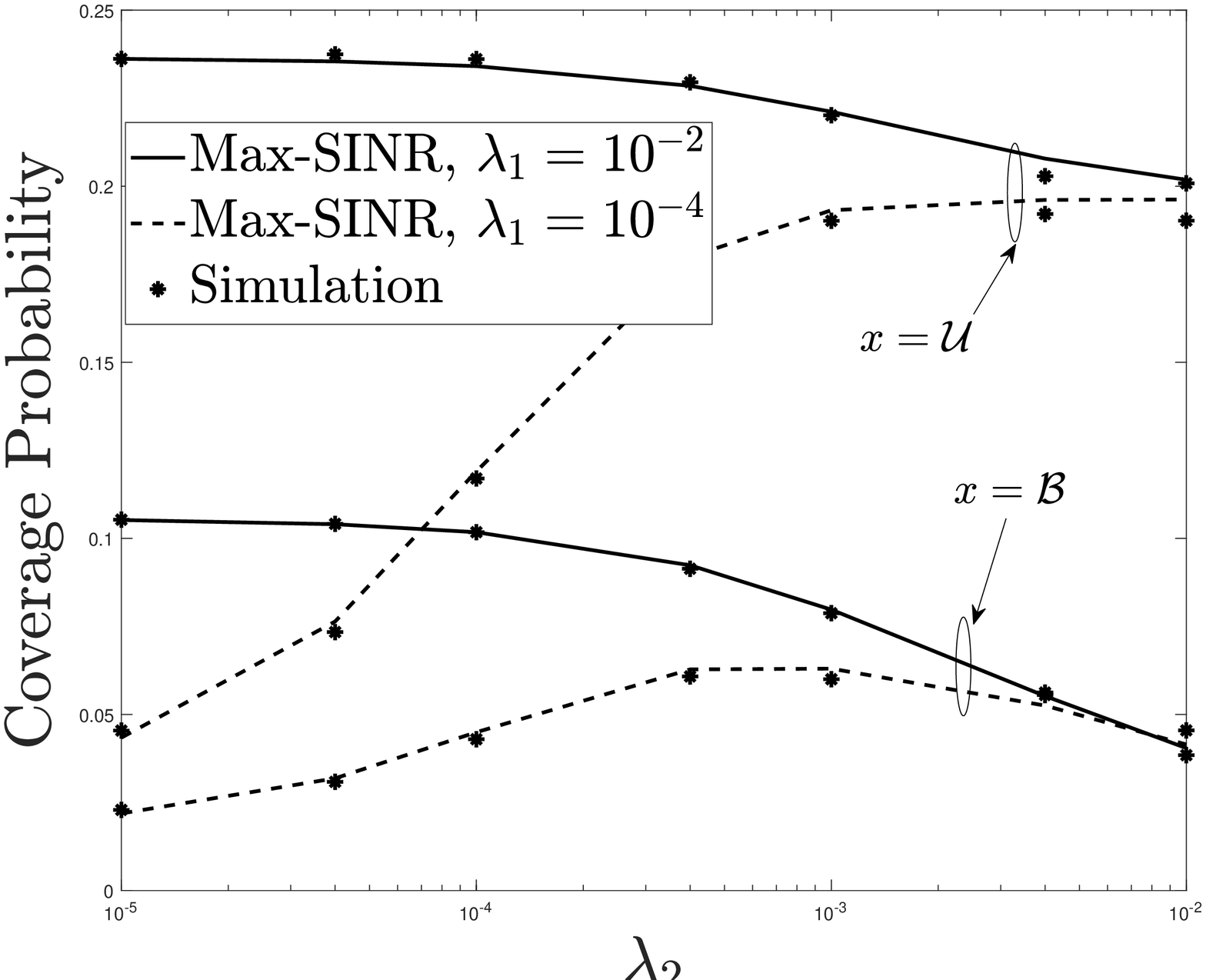}
\caption{Average coverage probability ${\cal C}^{ x}$  vs. $\lambda_2$  over $(3, 0, 0, 3)$-order  Fox's ${\mathcal H}$   for multipath fading  with ${\mathcal P}=(0.2, 5.5, -,(1.5, 0.4,4.5), - , \frac{1}{2}{\bf 1}_3)$. ${\cal M} = 2$, $\alpha=4$, $P_1 = 50$ W, $P_2 = 1$ W, and $\sigma_1^{2}=\sigma_2^{2}=10^{-6}$,  for: (a) both RSS and max-SINR CA schemes and $x =  \cal {U}$, and (b)  max-SINR CA and  $x \in  \{ \cal {U}, \cal {B}\}$..}
\end{figure}
\section{Conclusion}
Using a general form, namely the Fox's $\mathcal{H}$ variate of stochastic variables, we developed
a unifying  framework to characterize HetNet
communication under both RSS and Max-SINR CA rules. Our work systemises the use of the Fox's $\mathcal{H}$-function to incorporate prominent fading distributions and  bounded path-loss models.
We proposed generic closed-form expressions for the coverage
probability that reveal the actual impact of densification in conjunction with
the path-loss model, the fading parameters, and the SINR thresholds.

\section{Appendix A: Proof of Proposition 1}

\textit{ Definition 2 (Fox's $\he$  Transform \cite{mathai})}: The $\he$-transform of a
function $f(x)={\mathcal H}_{p_1,q_1}^{m_1,n_1}(x;{\mathcal P_1}=(\kappa_1,c_1, a_1, b_1,A_1,B_1))$
is defined by

\begin{eqnarray}
\!\!\!\!\!\!{\mathcal H}_{p,q}^{m,n}\left\{f(t); {\mathcal P}\right\}(s)\!\!\!&=&\!\!\!\!\int_{0}^{\infty} \!\!\!\!{\mathcal H}_{p,q}^{m,n}(t ;{\cal P}) f(t s) dt,\nonumber \\ \!\!\!&=&\!\!\!\!\frac{1}{s}{\mathcal H}_{p+q_1,q+p_1}^{m+n_1,n+m_1}(s^{-1}; {\mathcal P}\odot{\mathcal P_1}),
\label{trans}
\end{eqnarray}
where
\begin{eqnarray}
{\mathcal P}\odot{\mathcal P_1}\!\!\!\!\!&\triangleq&\!\!\!\!\!\!\bigg(\frac{{\mathcal\kappa}{\mathcal\kappa}_1}{c_1}, \frac{c}{c_1}, (1\!-\!b_1\!-\!B_1, a), (b^{1:m}, 1-a_1-A_1, \nonumber\\ && \!\!\!\!b^{m+1:q}),(B_1,A), (B^{1:m}, A_1,B ^{m+1:q})\!\bigg).
\end{eqnarray}
\textit{Proof:}  Follows from the Mellin transform of the product
of two $\he$-functions \cite[Eq. (2.3)]{mathai}.

Resorting to \cite[Theorem 1]{trigui} and \cite[Eq. (39)]{trigui1} under the independency of $\{\Phi_j\}$ and then applying the Fox's $\mathcal{H}$-transform in (\ref{trans}), we have
\begin{eqnarray}
{\cal C}^{\cal U}(r_k) &=& \int_{0}^{\infty}\frac{1}{\sqrt{\xi}}~ {\cal L}^{-1}\left\{\frac{1}{\sqrt{s}}{\mathcal H}_{p,q}^{m,n}\left\{f(t); {\mathcal P}\right\}(s \xi); s; {\cal \beta}_k\right\} \nonumber\\ &&e^{-\sigma^{2}_k\xi \frac{r_k^{\alpha}}{P_k}}
\prod_{j\in {\cal M}}{\cal L}_{{\cal I}_j}\left(\xi\frac{r_k^{\alpha}}{P_k}\right) d\xi,
\label{p1}
\end{eqnarray}
where $f(t)=\sqrt{t}{\mathcal J}_1\left(2\sqrt{s t \xi }\right)$,   ${\mathcal J}_{1}(x)={\mathcal H}_{0,2}^{1,0}\left(\frac{x^{2}}{4};(1,1, \frac{1}{2},-\frac{1}{2},1,1 )\right)$ is the Bessel function of the first kind \cite[Eq. (8.402)]{grad}, and ${\cal L}^{-1}$ is the inverse Laplace transform.
Moreover in (\ref{p1}), ${\cal L}_{{\cal I}_j}$ is the Laplace transform of the aggregate interference from the $j$-th tier  evaluated as in \cite[Eq. (43)]{trigui1}  as
\begin{equation}
{\cal L}_{{\cal I}_j}(\xi)\! = \!\exp \left(\!-\pi \delta \lambda_j \frac{\xi r_k^{2-\alpha}}{\left(1-\delta\right)}{\mathcal H}_{p,q}^{m,n}\left\{g(t); {\mathcal P}\right\}(\xi)\!\right),
\label{IL}
\end{equation}
where $g(t)=t ~{\rm {}_{2}F_{2}}\left(1, 1-\delta; 2;2-\delta; - \xi t r^{-\alpha}\right)=t{\mathcal H}_{2,3}^{1,2}\left(t;{\cal P}_1\right)$, ${\cal P}_1=(1-\delta, \xi r^{-\alpha}, (0,\delta), (0,-1,\delta-1), {\bf 1}_2, {\bf 1}_3 )$, and ${\rm {}_{p}F_{q}}(\cdot)$ is the generalized  hypergeometric function of \cite[Eq. (9.14.1)]{grad}.
Finally, applying \cite[Eq. (1.58)]{mathai}, the ${\mathcal H}$-transform in (\ref{trans}) and the inverse Laplace transform of the Fox's $\mathcal{H}$-function  \cite[Eq. (2.21)]{mathai} given by
\begin{equation}
{\cal L}^{-1}\{x^{-\rho}{\mathcal H}_{p,q}^{m,n}(x;{\cal P}); x;t\}=t^{-\rho-1}{\mathcal H}_{p+1,q}^{m,n}\left(\frac{1}{t};{\cal P}_l\right),
\end{equation}
where ${\cal P}_l=(\kappa, c, (a,\rho), b, (A,1), B)$,  the desired result is obtained  after applying the Fox's $\he$ reduction formulae in \cite[Eq. (1.57)]{mathai}.
The  coverage probability over Fox's $\mathcal{H}$-fading\!\footnote{We dropped the index $i$ from Fox's $\mathcal{H}$-distribution $\{{\cal O}_i, {\mathcal P}_i\}$ for notation simplicity.} with unbounded path-loss model for a receiver connecting to a $k$-th tier BS located at $x_k$ is given by
\begin{eqnarray}
{\cal C}^{\cal U}(r_k)\!\!&=&\!\!\int_{0}^{\infty}\frac{1}{\xi^{2}}{\mathcal H}_{q,p+1}^{n,m}\left(\xi; {\cal P}^{k}_{\cal U}\right)\exp\Bigg(-\frac{\sigma^{2}_k}{P_k}\xi r_k^{\alpha}\nonumber\\ &&\!\!\!\!\! \!\!\!\!\!\!\!- \pi \delta \sum_{j\in {\cal M}}r_k^{2} \lambda_j \widetilde{P}_j^{\delta} \xi {\mathcal H}_{q+2,p+3}^{n+1,m+2}\left(\xi; {\cal P}^{{\cal I}}_{\cal U}\right)\Bigg)d\xi,
\label{cux}
\end{eqnarray}
where $\widetilde{P}_j=\frac{P_j}{P_k}$, $\delta=\frac{2}{\alpha}$,  and the parameter sequences  ${\cal P}^{k}_{\cal U}=\!\!\left(\kappa {\cal \beta}_k, \frac{1}{c {\cal \beta}_k}, 1\!-\!b, (1\!-\!a, 1), \mathcal{B}, (A,1)\right),
$ and ${\cal P}^{{\cal I}}_{\cal U}=\bigg(\frac{\kappa}{c^{2}}, \frac{1}{c}, (1\!-\!b\!-\!2B,0,\delta),(0, 1\!-\!a\!-\!2A, -1,\delta\!-\!1),$ $ (\mathcal{B},1,1), (1,A,1,1)\bigg)$.
 Recall under the RSS CA  that the  PDF of the link's distance $r_k$ in HetNets is given by $f_{r_k}(x)=\frac{2\pi  \lambda_k}{\theta_k}  x \exp\left(-\sum_{j\in {\cal M}}\pi x^{2} \lambda_j \widetilde{P}_j^{\delta}\right)$ \cite{and3}. Then recognizing that $\exp(-x)={\mathcal H}_{0,1}^{1,0}(x;{1,1, 0, 1, 1,1})$ \cite[Eq. (1.125)]{mathai} in (\ref{cux}), we apply (\ref{trans}) to obtain the average coverage probability  in (\ref{cb}) after some manipulations.

\section{Appendix B: Proof of Proposition 2}
The proof of this Proposition relies on the very same approach adopted in Appendix A, yielding
\begin{eqnarray}
{\cal C}^{\cal B}(r_k)\!\!\!\!&=&\!\!\!\! \int_{0}^{\infty}\frac{1}{\sqrt{\xi}} ~ {\cal L}^{-1}\left\{\frac{1}{\sqrt{s}}{\mathcal H}_{p,q}^{m,n}\left\{f(t); {\mathcal P}\right\}(s \xi); s; {\cal \beta}_k\right\}\nonumber\\ && e^{-\sigma^{2}_k\xi \frac{(1+r_k)^{\alpha}}{P_k}}
\prod_{j\in {\cal M}}{\cal L}_{{\cal I}_j}\left(\xi\frac{(1+r_k)^{\alpha}}{P_k}\right) d\xi,
\label{p1}
\end{eqnarray}
where rearranging \cite[Eq. (39)]{trigui1} after carrying out the  change of variable relabeling $(1+x)^{-\alpha}$ as $x$, we have
\begin{eqnarray}
\!{\cal L}_{{\cal I}_j}(\xi)\!\!\!\! &=& \!\!\!\!\exp \!\!\Bigg(\!\!\!-\pi \delta \lambda_j \xi \Bigg(\!\!\frac{(1\!+\!r_k)^{2-\alpha}}{\left(1-\delta\right)}{\mathcal H}_{p,q}^{m,n}\!\!\left\{g_1(t); {\mathcal P_1}\right\}(\xi)-\nonumber\\ &&\frac{(1+r_k)^{1-\alpha}}{\left(1-\frac{\delta}{2}\right)}{\mathcal H}_{p,q}^{m,n}\left\{g_2(t); {\mathcal P}\right\}(\xi)\Bigg)\Bigg),
\label{IL}
\end{eqnarray}
where $g_1(t)=t ~{\rm {}_{2}F_{2}}\left(1, 1-\delta; 2; 2-\delta; - \xi t (1+r_k)^{-\alpha}\right)$ and $g_{2}(t)=t ~{\rm {}_{2}F_{2}}\left(1, 1-\frac{\delta}{2}; 2;2-\frac{\delta}{2}; - \xi t (1+r_k)^{-\alpha}\right)$.
Finally  applying (\ref{trans}) and  plugging the obtained result back
into (\ref{p1}), Proposition 2 then follows after some manipulations.

\section{Appendix C: Proof of Proposition 3}
Referring to \cite{trigui1},  the Laplace transform of the ICI from tier
$j$ under max-SINR CA is evaluated as ${\cal L}_{{\cal I}_j}(\xi)= \exp\left(- \pi  \lambda_j \xi^{\delta}\Gamma\left(1-\delta\right)  {\cal E}[H^{\delta}]\right)$, where
${\cal E}[H^{\delta}]$ is the Mellin transform of the Fox's-${\mathcal H}$ function obtained as  ${\cal E}[H^{\delta}]=\Lambda^{m,n}_{p,q}$ \cite[Eq. (2.8)]{mathai}. Then following the same lines developed in Appendix A yields
\begin{eqnarray}
{\cal C}^{\cal U}(r_k)&=&  \int_{0}^{\infty}\frac{1 }{\xi^{2}}{\mathcal H}_{q,p+1}^{n,m}\left(\xi; {\cal P}^{k}_{\cal U}\right)\exp\Bigg(-\frac{\sigma^{2}_k}{P_k}\xi r_k^{\alpha}\nonumber\\ &&\!\!\!\!\! \!\!\!\!\!\!\!-\sum_{j\in {\cal M}}r_k^{2} \pi \lambda_j \widetilde{P}^{\delta}_j \left(\frac{\xi}{c}\right)^{\delta}\Gamma(1-\delta)\Lambda^{m,n}_{p,q} \Bigg)d\xi.
\label{cm1}
\end{eqnarray}
Finally, substituting (\ref{cm1}) into (\ref{cm}) and applying (\ref{trans}) along with \cite[Eq. (1.59)]{mathai} yield
\begin{eqnarray}
\!\!\!{\cal C}^{\cal U}\!\!\!\!&=& \!\!\!\!\! \sum_{k\in {\cal M}}\lambda_k {\mathcal H}_{1,1}^{1,1}\left(1;\tilde{{\cal P}}_k \right)\int_{0}^{\infty}\frac{1 }{\xi^{\delta+2}}{\mathcal H}_{q,p+1}^{n,m}\left(\xi; {\cal P}^{k}_{\cal U}\right)
 d\xi \nonumber \\
&=& \sum_{k\in {\cal M}}\frac{\lambda_k }{  \beta_k^{\delta}}~{\mathcal H}_{1,1}^{1,1}\left(1;\tilde{{\cal P}}_k \right)\widetilde{\Lambda}^{n,m}_{q,p+1},
\label{out}
\end{eqnarray}
 where
\begin{eqnarray}
\widetilde{\Lambda}^{n,m}_{q,p+1}&=&\frac{\kappa}{c^{\delta+1}\Gamma(2+\delta)}\nonumber \\ &&  \!\!\!\!\!  \!\!\!\!\!  \!\!\!\!\!  \!\!\!\!\!  \!\!\!\!\!  \!\!\!\!\! \frac{\prod_{j=1}^{n}\Gamma\left(1-a_j-(1+\delta)A_j\right)
\prod_{j=1}^{m}\Gamma\left(b_j+(1+\delta)B_j\right)}{\prod_{j=n+1}^{p}\left(a_j+(1+\delta)A_j\right)\prod_{j=m+1}^{q}\Gamma\left(1-b_j-(1+\delta)A_j\right)}\nonumber \\ &=&\frac{\Lambda^{n,m}_{p,q}}{\Gamma(1+\delta)}.
  \end{eqnarray}

\clearpage

\begin{figure}[t]
\centering
\includegraphics[scale=0.8]{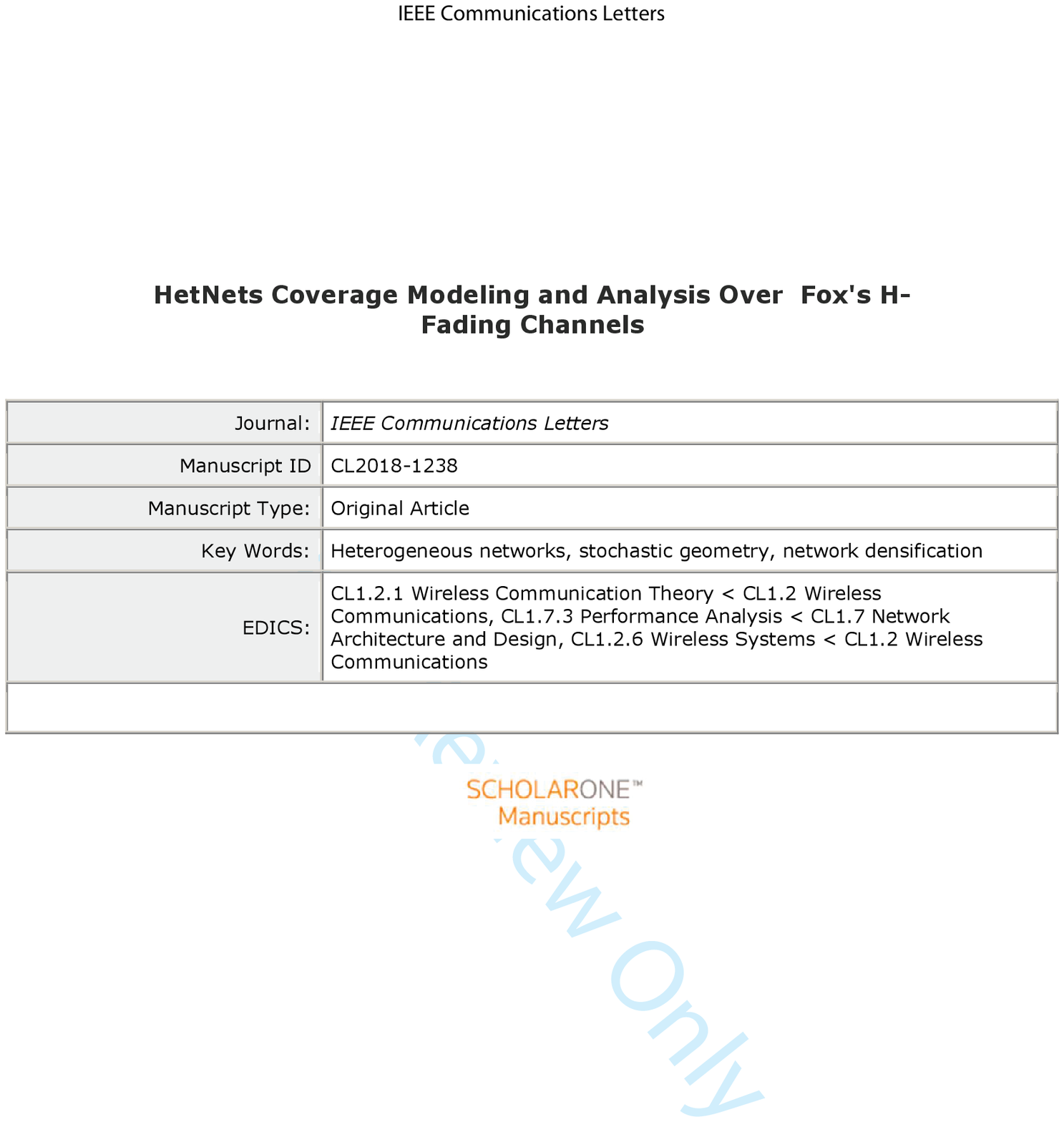}
\caption{Central Manuscript Submission Cover Page.}
\end{figure}

 \end{document}